\documentclass[12pt]{article}
\usepackage{graphicx,a4} 
\usepackage{amssymb}
\begin{document}

\newcommand{\preprintno}[1]
{\vspace{-2cm}{\normalsize\begin{flushright}#1\end{flushright}}\vspace{1cm}}

\title{\preprintno{{hep-ph/0112279}\\{\bf MCTP-01-65}\\{\bf ULB-TH/01-40}} 
Time-varying coupling strengths, nuclear forces and unification}
\author{Thomas Dent\thanks{email: tdent@umich.edu}\\
{\em Michigan Center for Theoretical Physics, Randall Lab.,}\\
{\em University of Michigan, Ann Arbor, MI 48109-1120 U.S.A.}\\
\\
Malcolm Fairbairn\thanks{email: mfairbai@ulb.ac.be}\\
{\em Service de Physique Th\'eorique, CP225}\\
{\em Universit\'e Libre de Bruxelles, B-1050 Brussels, Belgium}}

\date{July 2002}

\maketitle
\begin{abstract}
\noindent
We investigate the dependence of the nucleon-nucleon force in the deuteron
system on the values of coupling strengths at high energy, which will in 
general depend on the geometry of extra dimensions. The stability of 
deuterium at all times after nucleosynthesis sets a bound on the time 
variation of the ratio of the QCD confinement scale to light quark masses. 
We find the relation between this ratio, which is exponentially sensitive 
to high-energy couplings, and fundamental parameters, in various classes 
of unified theory. Model-dependent effects in the Higgs and fermion mass 
sector may dominate even over the strong dependence of the QCD scale $\Lambda$. 
The binding energy of the deuteron also has an important effect on 
nucleosynthesis: we estimate the resulting bounds on variation of couplings.
\end{abstract}

\section{Introduction}
In many models of particle physics, the universe is assumed to have more than 4 dimensions. The extra dimensions are either compactified to such a small size that we cannot (currently) probe them experimentally \cite{polchinski} or possess metrics with a nontrivial dependence on the transverse directions such that we can only detect the gravitational influence of our familiar 4 dimensions \cite{randall}. Cosmological solutions of the field equations of these theories often involve time evolution of the higher dimensions; the value of the gauge couplings in the low energy limit of these theories is invariably a function of the size or shape of the higher dimensions. This can be extremely problematic as variation in the gauge couplings over cosmological time scales may destroy the successful predictions of primordial nucleosynthesis \cite{olive}. Recent claims of a time-variation of the fine structure constant \cite{webb} also motivate the study of theories with dynamically-determined (and thus potentially time-varying) couplings.

In relation to nucleosynthesis, it is only clear what the effect of changing one coupling constant at a time has upon the light element abundance. It is possible that degeneracies in the effect on nucleosynthesis may arise when more than one gauge coupling changes at once: an overall change in all couplings might be acceptable at a level much greater than that permitted for any one on its own. In this situation it is not possible accurately to constrain models such as \cite{brax} where the gauge couplings oscillate with a fractional change of the order of 10\% in the matter dominated era. Such a large fractional change would not be acceptable in, for example, the electromagnetic fine structure constant $\alpha$ at nucleosynthesis, if this were the only time-dependent coupling. This paper is an attempt to provide an additional constraint, which is independent of nucleosynthesis (but which may affect calculations of nucleosynthesis), which suffers as little as possible from the problem of relating nuclear forces to underlying theory, and which is sensitive to a well-defined combination of couplings.

Deuterium is only produced during nucleosynthesis, as it is too weakly bound to survive in the regions of stars where nuclear processes take place. The fact that deuterium is still observed today means that variations in the gauge coupling strengths or other fundamental parameters are non-trivially constrained by the requirement that the deuteron be stable at all times after nucleosynthesis. The fact that the deuteron is so weakly bound also makes it more sensitive to variations in the internuclear force. The strong running of $\alpha_3$ at low energies means that a change in the coupling strength at high energy is manifested in a change in the strong coupling scale $\Lambda_{\rm QCD}$ (henceforth denoted by $\Lambda$), by the usual dimensional transmutation arguments. Changes in $\Lambda$ in turn lead to changes in the internuclear force. 

As recently pointed out by Langacker {\em et al.}\/\ \cite{LSS}, one also 
expects changes in quark masses and in the Higgs v.e.v.\ $v$ if gauge 
couplings are unified at some scale. Any viable unified theory should 
accommodate (if not predict) a mechanism for electroweak symmetry-breaking, 
which may well depend sensitively on SUSY-breaking masses, and a mechanism 
of generating small Yukawa couplings, all of which may have a dependence on 
the unified coupling. Moreover, $\Lambda$ is sensitive to all coloured 
particle masses through threshold effects of RG running. These effects 
introduce a large measure of model-dependence, since the correct theories 
of SUSY-breaking, electroweak symmetry-breaking and fermion masses are unknown.
One may choose for simplicity to set to zero unknown effects in the electroweak, 
SUSY and Yukawa sectors \cite{CalmetF}, but this runs the risk of neglecting
terms which are of equal size or larger than the terms kept in the analysis.

One might also consider ``less unified'' models, with more than one 
dynamically-determined fundamental coupling. The heterotic string dilaton
$S$ and volume moduli $T$ provide a basic example, where gauge couplings
and renormalisable Yukawa couplings (for canonically-normalised fields) have 
a universal dependence on $S$, but may be differently affected by changing 
$T$. The greater the number of independent quantities considered as 
time-dependent, the less predictive the theory becomes and the less meaningful
are any constraints. Here we restrict ourselves to estimating the dependence 
of low-energy quantities in a somewhat idealised framework with a single 
dynamical unified coupling, corresponding to the v.e.v.\ of a dilaton-like field.

In the first part of this paper we calculate the deuteron binding energy by
considering meson exchange forces, expressing the relevant parameters as a 
functions of $\Lambda$ and the light quark masses $m_u$, $m_d$. 
The result is rather simple: we find that the deuteron is stable as long as 
the ratio $m_q/\Lambda$ is greater than a certain value, where $m_q=m_u+m_d$. 
We perform similar calculations for the dineutron and diproton systems in 
the same isospin multiplet and investigate the criterion for their 
stability. Then we relate $\Lambda$ and the quark masses to the QCD coupling 
strength and running masses at high energy using renormalisation group (RG) 
evolution. We take two cases, supersymmetric models with RG running up to the 
GUT scale (similar results will be obtained in the case of power-law unification 
in large extra dimensions), and nonsupersymmetric low-scale models with RG 
running up to a scale of a few TeV. The main result of this section is the 
exponential dependence of $\Lambda$ on the perturbative strong coupling 
$\alpha_3$ at high energy. We also find how the Higgs v.e.v.\ and SUSY-breaking 
masses influence the low-energy parameters.

Finally we consider how the bounds deduced from the two-nucleon system apply
to various types of high-energy model, in which $\alpha_3$ and the quark 
masses depend on model parameters (in particular the sizes of extra 
dimensions) which may be time-dependent. Thus, bounds on the possible
cosmological evolution of such models since nucleosynthesis can be obtained. 
The constraints from the dinucleon system will in general apply to a different 
combination of theory parameters from those arising from nucleosynthesis --- 
taking the two together bounds the variation of fundamental parameters in two 
directions. 
(In addition, there are many other observational bounds applying at various
much later epochs, discussed for example in \cite{CalmetF}, including some from 
direct laboratory measurement.)
We also point out for the first time and estimate the effect of changing the
deuteron binding energy on the process of helium formation at nucleosynthesis, 
which may give rise to stronger bounds.

\subsection{Relation to recent work}
The relation between a time-varying fine structure constant and other
observables in particle physics was also treated in \cite{BanksDD,CalmetF,LSS}.
%Banks {\em et al.}\/\ \cite{BanksDD}, Calmet and Frizsch \cite{CalmetF}
%and Langacker {\em et al.}\/\ \cite{LSS} have also worked on the 
%relationship between time-variation of the fine structure constant and 
%other observables in the context of unified field theories. 
In \cite{BanksDD} the resulting variation in the vacuum energy $V_0$ was 
estimated from general principles of QFT and found to be enormously larger 
than the cosmological bounds; hence, the authors concluded that a large 
number of implausibly accurate fine-tunings would be necessary for a time 
variation of the size that has been claimed to be consistent with field 
theory and cosmology. 
Such an argument is rather weak since it assumes, crucially, that the
cosmological constant problem is ``solved'' at the present time by
cancellations of different field theory contributions, all of which are
many orders of magnitude larger than the measured value of $V_0$. This can
hardly be a solid starting-point from which to set theoretical limits. The
alternative considered in \cite{BanksDD} was a self-tuning mechanism which
protects the four-dimensional spacetime were we live by dynamically
``absorbing'' the vacuum energy into the curvature of one or more extra
dimensions. Such a mechanism probably disallows inflation, but it is by no
means clear that it also rules out spontaneous symmetry-breaking as
claimed (the argument being that anything that ``cancels'' the vacuum
energy also removes the source for a scalar to roll to the minimum of its
potential). The existence of a dynamical time-scale for the self-tuning,
the role of thermal effects in creating an effective potential, and the
possibility that the extra-dimensional model may break four-dimensional
Lorentz invariance are possible ways out of the $D=4$ field theory
argument that $V(\phi)$ must vanish at all times (see \cite{Grojean} for
related discussions).

We prefer to take the majority point of view on the cosmological constant,
{\em i.e.} that it is our greatest theoretical area of ignorance and that
no credible way to explain its smallness currently exists, therefore very
little can be deduced from it, and certainly nothing related to quantum
effects in the theory (which are the main difficulty in accommodating
varying alpha). If we want to retain the semiclassical picture of matter
coupled to gravity, the only sensible interpretation is a very light
scalar, which interestingly would have about the same mass as
quintessence; due to the coupling to electromagnetism, such a scalar would
mediate composition-dependent forces which might have experimental
signatures \cite{DvaliZ}. But the vanishingly small scalar mass (even
if experimentally confirmed) would remain a mystery, in the absence of an
underlying theory which would explain why spacetime curvature was
apparently so insensitive to a cosmologically-evolving field theory.

Calmet and Fritszch \cite{CalmetF} calculate some of the consequences for 
low-energy physics of changing $\alpha$, within a GUT-like theory which
constrains the SM gauge couplings to be equal at a particular energy scale
(or at least to satisfy some fixed relation). They consider exclusively 
the effects on the strong interactions, with the unstated assumption that the 
mechanisms of electroweak symmetry-breaking, supersymmetry-breaking and fermion 
mass generation are held constant despite varying the unified coupling, hence 
that the quark and lepton masses, as well as $W$ and $Z$ masses, remain unchanged.
Our calculation of the dependence of $\Lambda$ and $M_N$ on $\alpha_3(\mu>m_t)$ is 
essentially identical to theirs, except for including the full dependence on 
thresholds. As noted above, such a study can only be a first step since the 
electroweak mass scales may also depend sensitively on the unified coupling
(for example in models of hidden sector SUSY-breaking, radiative electroweak 
symmetry-breaking and anomalous U$(1)$ flavour models).
%the nucleon mass, 
%which depends sensitively on the unified coupling through the strong running
%of the QCD coupling, and also mention the Fermi coupling. Our calculation of 
%the dependence of $\Lambda$ and $M_N$ on $\alpha_3(\mu>m_t)$ is 
%essentially identical to theirs, except that we treat explicitly the full 
%dependence on thresholds.  
%This seems rather unrealistic, 
%since in many unified models the mechanisms determining these masses do 
%depend on the gauge coupling (for example hidden sector SUSY-breaking, 
%radiative electroweak symmetry-breaking and anomalous U$(1)$ flavour 
%models). 
%In fact the assumptions of their estimate of the variation in 
%$G_F$ appear somewhat doubtful, since (referring to the formulae $G_F 
%\propto g_2^2/M_W^2$ and $M_W\propto g_2 v$) one cannot simultaneously 
%vary $g_2$ and keep $M_W$ constant without varying $v$, but such a 
%variation will inevitably cause fermion masses to change. 

Unfortunately, since we do not know the correct theories of electroweak and 
supersymmetry-breaking, let alone that of fermion masses, correlations between 
the time-variation of different low-energy observables (such as $\alpha$ and the 
proton magnetic moment $\mu_p$ or the mass ratio $m_p/m_e$) are necessarily 
model-dependent, unless one can find a combination insensitive to, say, 
the electroweak sector. We take the converse approach and quote in a hopefully 
less model-dependent way the bounds on high-energy parameters deriving from 
the low-energy system that we are studying. In particular one cannot claim yet 
(as in \cite{CalmetF}) that the current data on the time variation of $\alpha$ 
and other quantities are inconsistent with unification; but given an 
observational bound one can derive bounds on time-varying fundamental parameters, 
in whatever model one is interested in. 

With precise measurements of at least two different quantities at any particular 
epoch, of which at least one shows a nonzero variation, one {\em can}\/ rule out 
classes of unified theory that predict the wrong relation between (variations 
in the) different quantities. Time-varying couplings in principle are a new way 
of doing phenomenology, which allow us to test relations between the {\em 
derivatives}\/ of different quantities rather than their static values.
This paper presents an additional bound which applies at all times after 
nucleosynthesis, thus for models in which time variation was more rapid in 
the early Universe (such as \cite{brax}) it is likely quite restrictive.

Langacker {\em et al.}\/\ \cite{LSS} have a somewhat similar approach to the 
``high-energy'' aspect of the problem, parameterising the effects of variations 
in the unified coupling strength on the electroweak and Yukawa sectors by 
unknown, model-dependent constants of proportionality which depend on the 
particular model used. By looking at other precision astronomical measurements      besides \cite{webb}, they find that these constants must satisfy a rather 
precise relation, for the model to be consistent with observation. 
%We note that \cite{LSS} and 
%\cite{CalmetF} differ substantially in their estimates of the change in the 
%magnetic moment $g$-factors. The relation $\Delta y/y \approx -121 
%\Delta \alpha/\alpha$, where $y\equiv \alpha^2 g_p$, appears in \cite{CalmetF},
%with the major contribution supposed to arise from the variation of $\Lambda$, 
%while in \cite{LSS} one finds $\Delta X/X \approx -32 \Delta \alpha/\alpha$, 
%where $X\equiv \alpha^2 g_p m_e/m_p$, with the variation in $g_p$ taken to
%be {\em negligible}\/ (since it arises from Clebsch factors in the constituent
%quark model) and the variation in $\Lambda$ dominating via the proton mass. 
%Even ignoring the difference in the estimates for $g_p$, the relation between
%$\Delta \Lambda/\Lambda$ and $\Delta \alpha/\alpha$ is apparently inconsistent
%between the two works. 

\section{Nuclear forces and the stability of di-nucleons}

\subsection{Chiral symmetry breaking and the pion} \label{sec:pi}
Previous calculations taking into account the effect of the time variation of the strong interaction on low energy nuclear phenomena were performed in the chiral limit, {\em i.e.}\/\ the limit of massless quarks and pions, where the only energy scale of the system is $\Lambda$. The approach of \cite{sis1} was to assume that certain dimensionful static quantities (for instance vector meson masses and nuclear binding energies) were directly proportional to $\Lambda$; in this way it is possible to place constraints on the variation of $\alpha_{3}$ over time. 

Explicit breaking of chiral symmetry in QCD is achieved by the addition of non-zero quark masses to the Lagrangian. The spontaneous breaking of chiral symmetry occurs when the quark condensate $\langle 0|q\bar{q}|0\rangle$ develops a non-zero value and dynamically lines up with the quark masses in the internal SU$(n_f)$ flavour space \cite{weinberg2}. The pion is the Goldstone boson associated with this spontaneous symmetry breaking: it is not quite massless, since the chiral symmetry is explicitly broken by the quark masses. The pion is however much less massive than any other strongly-interacting particle, which means that the internuclear pion force is relatively long range and important in the analysis of the loosely bound deuteron and unbound di-neutron and di-proton. 

Models of nuclear structure utilising just the exchange of the $\omega$, $\sigma$ and $\rho$ mesons \cite{walecka} are quite successful. This is partly because the isospin dependence of the pion interaction is such that the average effect across an isospin symmetric nucleus consisting of many protons and neutrons cancels. However, no such cancellation occurs in two-nucleon systems, where the internuclear pion force is finite and well-defined. The large separation of the proton and neutron in the deuteron (and in the putative di-neutron and di-proton systems, if coupling strengths were to change so that they were weakly bound), relative to the typical range of internuclear forces, increases the importance of the long range pion force in the binding of the system. Consequently, the pion force is the dominant contribution to the binding of two-nucleon systems, and the contributions of $\omega$, $\sigma$ and $\rho$ meson exchange to the binding energy can be taken to be of secondary importance. 

The pion mass is given by the Gell-Mann-Oakes-Renner relation \cite{gell}
\begin{equation}
m_{\pi}^{2}f_{\pi}^{2}=(m_{u}+m_{d}) \langle 0|q\bar{q}|0\rangle
\label{pimass}
\end{equation}
The non-zero pion mass leads to a finite divergence of the total axial current, which compensates for the partially conserved axial current of the weak interaction Hamiltonian. This compensation leads to the Goldberger-Treiman relation for the pion-nucleon coupling $g_{\pi}$ \cite{gold}
\begin{equation}
g_{\pi}=\frac{2M_{N}g_{A}}{f_{\pi}}
\label{gold}
\end{equation}
where $M_N$ is the nucleon mass. Although this is an approximate relation, one would expect it to remain valid as the parameters $f_{\pi}$ and $M_N$ change, so long as the variation is much less than that necessary to restore the chiral symmetry of the vacuum at zero temperature. (If this were not the case, the problems created for nuclear physics in the early Universe would be much greater than those we discuss here, indeed such a scenario would be immediately ruled out).

Since the nucleon mass $M_N$ originates mainly from confinement of the quark colour charges (rather than current quark masses), $M_N$ simply scales in direct proportion to $\Lambda$. Combining this with Eqs.~(\ref{pimass}) and (\ref{gold}) we find the relation
\begin{equation}
\frac{g_{\pi}^{2}m_{\pi}}{M_{N}} = \frac{g_\pi^3 (m_u+m_d)^{1/2} \langle 0|q\bar{q}|0\rangle^{1/2}}{2g_A M_N^2} \propto \frac{g_\pi^3 (m_u+m_d)^{1/2} \langle 0|q\bar{q}|0\rangle^{1/2}}{g_A \Lambda^2}
\end{equation}
%\begin{equation}
%\frac{g_{\pi}^{2}m_{\pi}}{M_{N}} =  
%\frac{\Lambda g_{A}^{2}(m_{u}+m_{d})^{\frac{1}{2}}{\langle}0|q\bar{q}|0\rangle^{\frac{1}{2}}}{f_{\pi}^{3}}
%\end{equation}
Note that the product $m_q \langle 0|q\bar{q}|0\rangle$ is invariant under 
change of renormalisation scale in QCD (see {\em e.g.}\/\ \cite{Jamin:2001zr}) 
so thus far we avoid the problem of having scale-dependent quantities in a 
context where mass scales may be time-dependent (see section \ref{sec:RGscale}).

%scale of 
%chiral symmetry-breaking, which is proportional to $f_{\pi}$. This can be deduced 
%indirectly from finite temperature lattice calculations, where simulations with 
%one \cite{kogut} and two quark flavours \cite{karsch} have shown that chiral 
%restoration and deconfinement occur at the same temperature.  
To proceed further, we need the relationship between 
$\langle 0|q\bar{q}|0\rangle$ and $\Lambda$, obtained by looking at the effective 
potential for the chiral symmetry-breaking, which shows that the energy scale 
associated with chiral symmetry restoration varies as the value of the condensate:
\begin{equation}
\langle 0|q\bar{q}|0\rangle^{\frac{1}{3}}\sim\Lambda,
\end{equation}
a result also predicted by sum rule calculations of the nucleon mass 
\cite{nmass}. As it stands this is not a well-defined relation, 
since the LHS is re\-normal\-is\-ation-scale dependent. To remedy this, we use the 
formalism of ``invariant quark masses'' described {\em e.g.}\/\ in 
\cite{Narison}: to one loop the equation for the running quark mass 
$\bar{m}_i(\mu)$ is solved by
\begin{equation} \label{qinvmass}
\bar{m}_i(\mu) = \hat{m}_i(-\beta_1\alpha_3(\mu)/\pi)^{-\gamma_1/\beta_1}
\end{equation}
where $\hat{m_i}$ is a RG invariant quantity, $\beta_1$ is the one-loop 
beta-function coefficient and $\gamma_1$ the one-loop anomalous dimension of 
quark masses, such that for three flavours $-\gamma_1/\beta_1 = -2/(-9/2) = 4/9$. 
Thus for scales below $m_c$ we have 
$\bar{m}_i(\mu) = \hat{m}_i \alpha_3(\mu)^{4/9}$, 
up to a constant universal factor which will
drop out of fractional changes $\Delta(\ln \hat{m})\equiv \Delta\hat{m}/\hat{m}$.
Then the bilinear order parameter must have the one-loop behaviour  
\begin{equation}
\langle 0|q\bar{q}|0\rangle (\mu) = (\mbox{RG invariant})\times 
\alpha_3(\mu)^{-4/9} \propto \Lambda^3 \alpha_3(\mu)^{-4/9}
\end{equation}
which gives the correct dependence on both the RG scale and (a possibly 
time-varying) $\Lambda$.
We can now write
\begin{equation}
\frac{g_{\pi}^{2}m_{\pi}}{M_{N}}= {\rm const.} \times g_{A}^{2}\left(\frac{\hat{m}_{u}+\hat{m}_{d}}{\Lambda}\right)^{\frac{1}{2}}
\end{equation}
in which all RG dependence cancels neatly
\footnote{Such a redefinition can be performed to any desired loop order.}. 
We will see in the following sections that the value of the ratio 
$(\hat{m}_q/\Lambda)^{1/2}$, which we will denote as $c$,
\begin{equation}
c^2 \equiv \frac{\hat{m}_q}{\Lambda}
\end{equation}
is the parameter controlling the effect of varying coupling strengths on nuclear physics phenomena dominated by pion exchange. Later we shall see how $c$ can be related to coupling strengths at high energy in some examples of unified models.

Many different estimates of the dependence on nuclear binding energies on the one
dimensionless parameter $c^2$ exist; despite the fact that the deuteron is
the simplest nuclear system, a level of understanding sufficient to
estimate $B_d$ reliably from first principles is lacking. A systematic
approach based on expanding in the light quark masses has been proposed
\cite{Beane01}, but even in this framework the lack of control of
four-nucleon operators introduces uncertainties in the estimation of $B_d$
(highlighted in \cite{Beane02}, which appeared after the first version of
this paper). Since we do not claim to be doing precision calculations of
$B_d$, and would be satisfied with an estimate of its dependence on $c^2$
which was inaccurate by a factor of a few (remember that we are placing
cosmological bounds on $\delta\alpha$), this is not a major concern.
However, Beane and Savage claimed in \cite{Beane02} that the dependence on
$c^2$ could be mostly or entirely erased for some values of coefficients
of these four-nucleon operators, drawing the conclusion that bounds on
varying couplings from nuclear physics were considerably weakened. Without
further analysis we cannot tell to what extent this particular choice of
coefficients is fine-tuned, but in the absence of an underlying reason it
appears very unlikely that the ultraviolet effects (from the point of view
of the chiral expansion) giving rise to the quark mass-dependence of
four-nucleon operators would conspire with the low-energy effects of pion
exchange in such a way. Since all previous estimates gave a rather steep
dependence of $B_d$ on $c^2$, such a conspiracy would be a very unlucky
(or lucky?) coincidence.

\subsection{The di-neutron}
In this section we attempt to calculate how large a change in the parameter $c$ would allow two neutrons to form a stable bound state. The total spin of the di-neutron ground state would be zero and the neutrons share the same orientation in isospin space so we use the potential for the $S=0$, $I=1$ state \cite{ericson}
\begin{equation}
V(r)_{(S=0,I=1)}=-\frac{f^2}{4\pi}\frac{e^{-m_{\pi}r}}{r}
\label{central}
\end{equation}
where the dimensionless coupling $f^2$ is given by
\begin{equation}
\frac{f^2}{4\pi}=\frac{g_{\pi}^2}{4\pi}\left(\frac{m_{\pi}}{2M_N}\right)^2=7.95\times 10^{-2}
\label{f}
\end{equation}
where the numerical value is derived from present-day measurements. We assume a trial wavefunction of the form
\begin{equation}
\psi(r)=e^{-1/m_{\pi}r}e^{-bm_{\pi}r}  
\label{wave}
\end{equation}
In this equation $\psi(r)\equiv r\Psi(r)$, where $\Psi$ is the radial part of the wavefunction, such that $\psi^2(r)dr$ is the probability of finding the nucleon separation to be between $r$ and $r+dr$. At large $r$ where the Yukawa potential is negligible, the Schr{\" o}dinger equation for the relative motion of the two nucleons gives 
\begin{equation}
\frac{1}{M_N}\frac{d^2\psi(r)}{dr^2}=-E_B \psi(r)
\label{bind}
\end{equation}
\begin{figure}[tb]
\centering
\includegraphics[width=10.5cm,height=8cm]{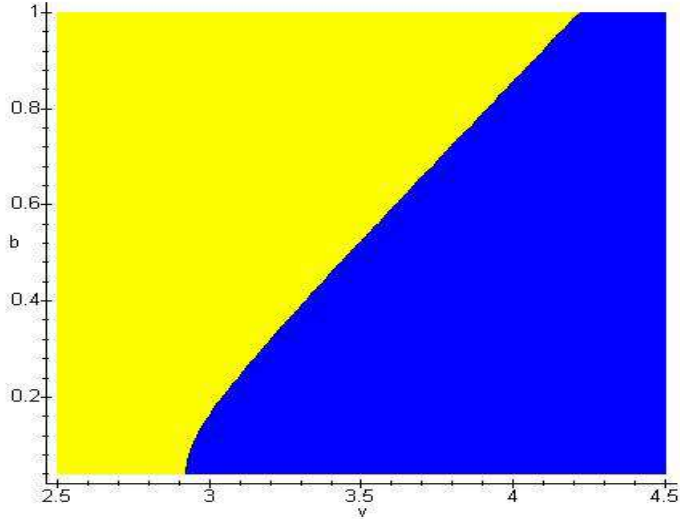}
\caption{Total ground state scaled energy of the di-neutron with respect to the parameters b and v introduced in the text. The light region corresponds to positive energy (unbound) and the dark region negative energy (bound).}
\label{contour}
\end{figure}
where $E_B$ is the (negative) binding energy of the state and the wavefunction (\ref{wave}) takes the form
\begin{equation}
\psi(r)=e^{-\sqrt{-E_B M_N}r},\qquad b\equiv\frac{\sqrt{-E_B M_N}}{m_{\pi}}.
\end{equation}
The trial wavefunction (\ref{wave}) is a suitable choice since at small $r$ it is independent of the binding energy whereas at high $r$ it is completely determined by the value $E_B$.
By applying the Hamiltonian to the wavefunction (\ref{wave}) we find the energy of the system \cite{wood}
\begin{eqnarray}
E&=&\frac{\int_{0}^{\infty} H(r) \psi(r) dr}{\int_{0}^{\infty} |\psi(r)|^{2} dr}\nonumber\\  
&=&-\frac{\sqrt{b}f^{2}m_{\pi}\mathcal{K}_0(2\sqrt{2+4b})}{4\pi\mathcal{K}_1(4\sqrt{b})}+\frac{b^{3/2}m_{\pi}^{2}\mathcal{K}_2(4\sqrt{b})}{2M_{N}\mathcal{K}_1(4\sqrt{b})}
\label{hamil}
\end{eqnarray}
where the $\mathcal{K}_n$ are modified Bessel functions of the second kind. We define a dimensionless parameter $v$ which is directly proportional to $c$ defined above
\begin{equation}
v=\frac{f^2}{4\pi}\frac{2M_N}{m_{\pi}}\propto c.
\end{equation}
Thus, we can express the scaled ground state energy, in units of $bm_{\pi}^2/2M_N\mathcal{K}_1(4\sqrt{b})$, in the simplified form
\begin{equation}  
E_{\rm scaled}=b\mathcal{K}_0\left(2\sqrt{2+4b}\right)-v\mathcal{K}_1(4).
\end{equation}
The dependence of $E_{\rm scaled}$ on $b$ and $v$ is illustrated in Fig.~\ref{contour}.  In order for a bound state to exist, we require $v\ge 2.8$, or
\begin{equation}
\frac{v}{2}=\frac{f^2}{4\pi}\frac{M_N}{m_{\pi}}=\frac{g_{\pi}^2}{4\pi}\frac{m_{\pi}}{M_N}>1.4
\end{equation}
We denote by $v_0$ the currently-observed value of $\frac{f^2}{4\pi}\frac{2M_N}{m_{\pi}}$, numerically equal to $1.08$. Then the di-neutron stability criterion is
\begin{equation}
E_B(nn)<0 \Rightarrow \frac{v}{v_0}=\frac{c}{c_0}\ge 2.6
\end{equation}
thus if $c\equiv\sqrt{\hat{m}_q/\Lambda}$ increases by a factor of 2.6 the di-neutron will become bound.

\subsection{The di-proton}
The fact that nuclear forces are independent of electromagnetic charge is illustrated by the result that the difference between the binding energies of H$^3$ and He$^3$ is well explained by the energy of Coulomb repulsion between the two protons in He$^3$ \cite{blatt}. One would expect the ground state of the di-proton to have the same nuclear quantum numbers as the ground state of the di-neutron. We can therefore see from the previous section that reducing the Coulomb repulsion to zero will not be enough to bind the di-proton: only a large variation in the strong force (more precisely, in the ratio $\hat{m}_q/\Lambda$) would achieve this. 

One would expect the size of a marginally bound di-neutron or di-proton to be of the same order as the effective range of the potential responsible for the binding. For a Yukawa potential like the one in Eq.~(\ref{central}) that we are using, the effective range is of order of the pion mass
\begin{equation}
r_{\rm eff}\sim\frac{1}{m_{\pi}}
\end{equation}
and the Coulomb repulsion energy at that range is approximately 1 MeV. Thus the parameters necessary to obtain a bound di-neutron with a binding energy of 1 MeV will be similar to those required to bind the di-proton at threshold. We assume here that the effect of varying $\alpha$ on the electromagnetic repulsion will be dwarfed by the corresponding change in $\hat{m}_q/\Lambda$. This assumption is shown to be justified consistent in the next section. Equation (\ref{bind}) and Fig.~\ref{contour} show that this corresponds to the condition on the parameter $v$ of
\begin{equation}
E_B(pp)<0 \Rightarrow \frac{v}{v_0}=\frac{c}{c_0}\ge 3.2
\end{equation}
so if $c$ increases by a factor of 3.2 the di-proton will become bound.

There also exists a hypothetical proton-neutron pure $L=0$ state which one would expect tro be stabilised for similar values of $c$ as the di-neutron.

\subsection{The deuteron}
The deuteron has total isospin $I$ is zero, so the spins of the two nucleons must be parallel by the Pauli exclusion principle. The $L=0$ wavefunction is unbound in the same way as the di-neutron and the di-proton, however the parallel nucleon spins mean that the $L=2$ wavefunction is bound by the tensor force and the ground state forms an admixture of these two wavefunctions. The exact form of the wavefunctions is given by the solution of a coupled pair of 2nd. order differential equations \cite{ericson}. To solve these equations explicitly for different values of the parameters $f$, $m_{\pi}$ and $M_N$ is highly nontrivial, so we will resort to a simple square well model of the kind used in \cite{barrownuc}.  

The analysis is then simply the bound state condition for a finite square well potential of depth $V_{sq}$ and width $a$, found (assuming $|E_B|\ll |V_{sq}|$, which is the case in the situations considered) by matching a sinusoidal solution for the wavefunction inside the well to an exponentially decaying mode outside. We assume $a=m_{\pi}^{-1}$ then the binding energy $E_B$ is then given by the solution of 
\begin{equation}
\cot\left[m_\pi^{-1}\sqrt{M_N(E_B-V_{sq})}\right]=-\sqrt{\frac{E_B}{V_{sq}-E_B}}.
\label{cot}
\end{equation}
Then, using the observed values of the pion mass and the binding energy $E_B=-2.226\,$MeV we find that the depth of the well is $V_{sq}=-66.15\,$MeV. Since the prefactor in the internuclear pion potential $f^2/4\pi$ is dimensionless we need to see how the quantities in this simple model scale with the parameter $c$ that we have been using to parameterise the variation in the underlying gauge couplings. The corresponding dimensionful `depth' $V_{y}$ of a Yukawa potential is related to equation (\ref{central}) by
\begin{eqnarray}
V(r)=V_{y}\frac{e^{-m_{\pi}r}}{m_{\pi}r}&=&\frac{f^2}{4\pi}\frac{e^{-m_{\pi}r}}{r}\nonumber\\
\frac{V_{y}}{m_{\pi}}&=&\frac{f^2}{4\pi}
\end{eqnarray}
so we can expect $V_{sq}$, the depth of our square well, to be directly proportional to $f^{2}m_{\pi}$. Making the substitution $\beta=E_B/V_{sq}$ equation (\ref{cot}) becomes
\begin{equation}
\cot\left[\sqrt{\frac{M_N V_{sq}}{m_{\pi}^2}\left(\beta-1)\right)}\right]=-\sqrt{\frac{\beta}{1-\beta}}
\label{cotcot}
\end{equation}
so the relevant combination of parameters is $M_N V_{sq}/m_{\pi}^2$ 
\begin{equation}
\frac{M_N V_{sq}}{m_{\pi}^2}\propto f^2\frac{M_N}{m_{\pi}}\propto c.
\end{equation}
Using Eq.~(\ref{cotcot}) we can see that the value of $c$ where the deuteron becomes unstable is
\begin{equation}
\frac{c_{unstable}}{c_o}=\frac{\pi^2}{4}\frac{m_{\pi}^2}{M_N |V_{sq}|}=0.77
\end{equation}
so a 13\% reduction in $c$ will give rise to an unbound deuteron.

\section{Running of gauge couplings and unification}
\subsection{GUT-like models}
The LEP precision measurements of gauge couplings suggest that the three standard model gauge groups become unified at some high energy $M_{U}\approx 2\times 10^{16}\,$GeV \cite{amaldi}. Thus any variation in the unified gauge coupling $\alpha_{U}(M_{U})$ leads to calculable changes in the low energy values of the gauge couplings. The dependence of low energy couplings on $\alpha_{U}$ is very similar in the case of large extra dimensions \cite{dienes} or non-factorisable geometries \cite{pomarol} where the unification scale is much lower. This is because extra contributions to the renormalisation group equations from the massive Kaluza-Klein modes of the theories change the running in such a way that the effective energy scale and coupling strength of unification stays the same from the point of view of the low energy 4D theory (see, however, $\cite{rossghil}$). In the following analysis we start with the simplest scenario of a ``grand desert'' between the weak and GUT scales with ${\mathcal N}=1$ supersymmetry. However, no matter what the matter content or other details, results of a similar order to those presented here seem unavoidable for any theory with gauge unification. 

The running of the gauge couplings $\alpha_{i}$, $i=1$, $2$, $3$, with the renormalisation scale $\mu$ is given at one loop by the expression 
\begin{equation}
\frac{d\alpha_{i}}{dt}=-\frac{b_{i}\alpha_{i}^{2}}{2\pi},\ t\equiv\ln(\mu/\mu_{0})
\label{run}
\end{equation}
where $\mu_0$ is an arbitrary, constant reference scale and the $b_{i}$ depend upon the gauge groups and matter representations transforming under each group. In SU(5) and SO(10) SUSY GUT's and a large class of string models the gauge couplings are unified at some high energy scale $M_{U}$:
\begin{equation}
\alpha_{3}(M_{U})=\alpha_{2}(M_{U})=(5/3)\alpha_{1}(M_{U})=\alpha_{U}(M_{U})
\end{equation}
where the normalisation factor $5/3$ derives from the fact that the U(1) hyper-charge gauge group must be embedded consistently in the unified group or string model. Other normalisations are possible, but are often inconsistent with unification 
\cite{dienes}.
At energies above the superpartner masses, the $b_{3}$ and $b_{2}$ of Eq.~(\ref{run}) are given by \cite{bailin}
\begin{equation}
b_{3}=9-2n_{G},\ b_{2}=6-2n_{G}-\frac{n_{H}}{2}
\label{beta}
\end{equation}
where $n_{H}$ is the number of Higgs doublets and $n_{G}$ is the number of fermion generations. Integration of Eq.~(\ref{run}) above the scale of super\-symmetry-breaking yields
\begin{equation}
\frac{M_{U}}{M}= \exp\left[\frac{4\pi}{6+n_{H}}\left(\frac{1}{\alpha_{2}(M)}-\frac{1}{\alpha_{3}(M)}\right)\right].
\end{equation}
Using the standard relation $\alpha_1 \cos ^2 \theta_W =\alpha_2 \sin ^2 \theta_W=\alpha$ and substituting the observed values of $\sin^{2}\theta_{W}$, $\alpha_{3}(M_{Z})$ and $\alpha(M_{Z})$ \cite{groom} in this equation, we find a unification energy $M_{U}\approx 10^{16}$ GeV and a unified coupling of $\alpha_{U}^{-1}\approx 25$. 

In order to find the effect on the strong interactions of changing $\alpha_{U}$ we will use the $b_{3}$ of Eq.~(\ref{beta}) to run $\alpha_{3}$ from $M_{U}$ down to low energy for different values of $\alpha_{U}$. Then we assume that all the superpartners of the standard model particles have the same mass $\tilde{m}$, which defines the scale below which supersymmetry is broken. At energies less than $\tilde{m}$ the running of $\alpha_{3}$ proceeds as in normal QCD, {\em i.e.}\/\
\begin{equation}
\frac{1}{\alpha_{3}(M_{\rm high})}-\frac{1}{\alpha_{3}(M_{\rm low})}=-\frac{33-2n_{f}}{6\pi}\ln\left(\frac{M_{\rm low}}{M_{\rm high}}\right)
\end{equation}
where $n_{f}$ is the number of quark flavours of mass $m_q < m_{\rm high}$. We use the observed values of the top, bottom and charm quark masses, defined at the self-consistent renormalisation scale $m_q(m_q)$, in this part of the running.

If one neglects quark masses, the Lagrangian of QCD contains no dimensionful parameters: however, on quantisation, loop corrections give rise to the renormalisation group invariant\footnote{Invariant unless one crosses the mass threshold of a quark or superpartner species, which changes the running and the apparent value of $\Lambda$. Here, we calculate $\Lambda$ under the assumption that it lies at the energy scale $m_s < \Lambda < m_c$.} strong interaction scale $\Lambda$: 
\begin{equation}
\Lambda = M {\rm{exp}} \left(\frac{-6\pi}{(33-2n_{f})\alpha_{3}(M)}\right)
\label{lambda}
\end{equation}
where $M<\tilde{m}$. This scale sets the characteristic energy of the particles of the low energy effective theory. Even in the presence of massive quarks, this quantity remains important in understanding low energy phenomena, as we saw in the calculation of internuclear forces.

For constant $\alpha_{U}$, the running of $\alpha_3$ below the superpartner 
thresholds and the value of $\Lambda/M_U$ depend strongly on the common 
superpartner mass $\tilde{m}$. But if $\tilde{m}/M_U$ is held fixed and 
$\alpha_{U}$ varied, the fractional change in $\Lambda/M_U$ for a given fractional change in $\alpha_{U}$ has little dependence on the value of $\tilde{m}$. For any constant value of $\tilde{m}/M_U$ correponding to superpartner masses between $100\,{\rm GeV}$ and $2\,{\rm TeV}$, we find the same ratio of fractional changes $(M_U/\Lambda)\Delta(\Lambda/M_U)/(\Delta\alpha_{U}/\alpha_{U})$, to well within the accuracy of our results. Later, we give the explicit formula for $\Lambda$ in the case $\tilde{m}/M_U$ is also changing.

\begin{figure}[tb]
\centering
\includegraphics[width=10.5cm,height=8cm]{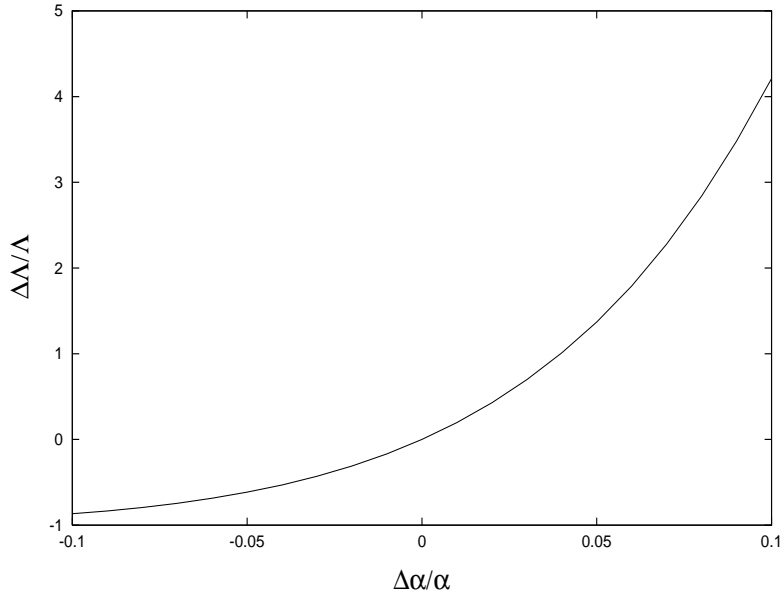}
\caption{Variation in $\Lambda$ (strictly, $\Lambda/M_U$) vs.\ variation in the gauge unified gauge coupling.}
\label{variation}
\end{figure}

In order to obtain $\alpha_{3}$ at low energies and $\Lambda/M_U$ as a function of 
$\alpha_{U}(M_{U})$ we include the masses of the top, bottom and charm quarks 
\cite{groom} and change the running accordingly, with results illustrated in 
Fig.~\ref{variation}.
To first order in the fractional change $\Delta\alpha_{U}/\alpha_{U}$, we
should be able verify the results for $\Delta(\ln \Lambda)$ in \cite{CalmetF,LSS}. 
Later, we will consider the effect of varying the threshold masses, but for the 
moment we consider them as fixed. Using the result quoted later as 
Eq.~\ref{eq:LambdaMSSM} we find 
\begin{equation}
\Delta(\ln \Lambda/M_U) = \frac{6\pi}{27} \alpha_U^{-1} \Delta(\ln \alpha_U)
\approx 17 \Delta (\ln \alpha_U)
\end{equation}
consistent with the result of both papers. (Note that any determination 
of the absolute value of $\Lambda/M_U$ will depend significantly on the 
renormalisation prescription, so without a detailed treatment, including also
higher loop effects, our estimate of the prefactor is likely rather imprecise.)

\subsection{Non-GUT (brane world-type) models}
\begin{figure}[tb]
\centering
\includegraphics[width=10.5cm,height=8cm]{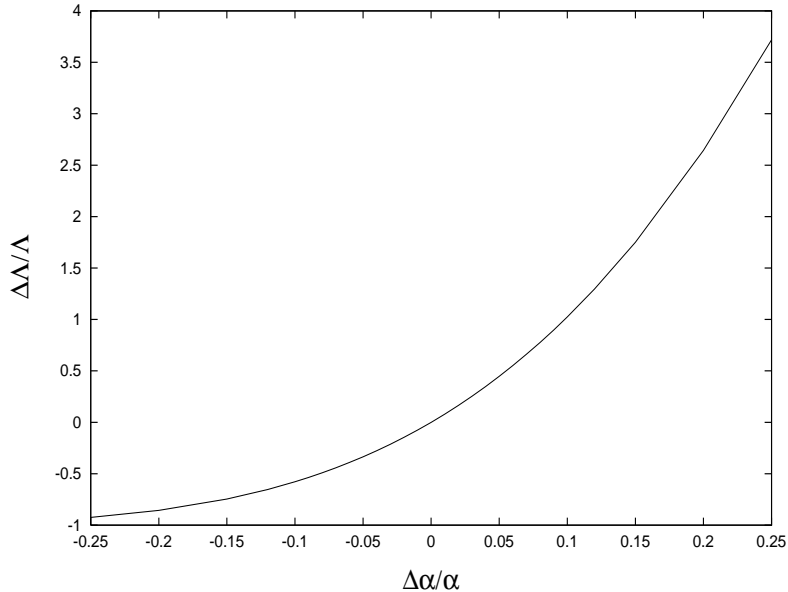} 
\caption{Variation in $\Lambda/M_*$ vs.\ variation in $\alpha_3$ at a scale of $1\,$TeV.}
\label{vartev}
\end{figure}

One can perform a similar analysis in the case of theories where the gauge couplings at the fundamental scale do not satisfy a GUT relation, for example intersecting brane models (see {\em e.g.}\/\ \cite{Cvetic,AldazFLRU}) where different gauge groups propagate along different world-volumes (\cite{berlin}). In this case we just find the relation between the SU$(3)$ coupling just below the fundamental scale $M_*$ and $\Lambda$, by the same method. For a high-scale model with softly-broken SUSY the calculation is identical, replacing $\alpha_{U}$ by $\alpha_3(\mu\simeq M_*)$ (where $M_*$ is the scale below which $d=4$ effective theory applies for gauge interactions). In the case of a low fundamental scale of order $1$-$10\,$TeV and no supersymmetry we simply run $\alpha_3$ from $M_*\sim {\rm TeV}$ down through the heavy quark thresholds, with results as illustrated in Fig.~\ref{vartev}.
The main lesson from this exercise is the sensitivity of $\Lambda/M$ to the 
strong coupling $\alpha_3$ at high energies $M\gg M_N$: the slope of both graphs 
around $\Delta=0$ is 10 to 20. In our conclusions, for simplicity and to 
avoid making highly model-dependent statements, we will pretend that the 
analysis finishes at this point, and express the results in terms of a 
fractional change in $\alpha_U$ under the (in general unrealistic) assumption of 
constant quark and superpartner masses. 

\section{Thresholds and RG running} \label{sec:thresh}
In both cases (GUT-like and non-GUT) the running of $\alpha_3$ is complicated 
by the dependence on heavy quark and coloured superpartner (gaugino and squark)
thresholds. In the case with superpartners the one-loop result with
tree-level matching at thresholds is
\begin{equation} \label{eq:LambdaMSSM}
\Lambda = M e^{-6\pi/27\alpha_3(M)} \left(\frac{m_c m_b m_t
m_\lambda^3 \overline{m_{\tilde{q}}}^3}{M^9}\right)^{2/27}
\end{equation}
where $m_\lambda$ is the gaugino mass, $\overline{m_{\tilde{q}}}$ is the
geometric average squark mass and $M$ is an arbitrary scale above the 
superpartner masses; for the SM we find simply
\begin{equation}
\Lambda = M' e^{-6\pi/27\alpha_3(M')} \left(\frac{m_c m_b m_t}
{{M'}^3}\right)^{2/27}
\end{equation} 
where $M'>m_t$. The quark and superpartner masses are the
``decoupling masses'' $m_q(\mu=m_q)$ {\em etc.} in a convenient
renormalisation scheme.

The possibility of varying threshold masses complicates the simple
one-loop formula
\[
\frac{1}{\alpha(\mu)}\frac{\Delta\alpha(\mu)}{\alpha(\mu)} =
\frac{1}{\alpha(\mu')}\frac{\Delta\alpha(\mu')}{\alpha(\mu')} \]
derived in \cite{CalmetF} for a fixed ratio $\mu/\mu'$ if there is, say, a
quark with $\mu<m_{q_i}<\mu'$; we find
\[
\frac{1}{\alpha(\mu)}\frac{\Delta\alpha(\mu)}{\alpha(\mu)} =
\frac{1}{\alpha(\mu')}\frac{\Delta\alpha(\mu')}{\alpha(\mu')} +
\frac{b^{>m_q}-b^{<m_q}}{2\pi} \Delta\ln \frac{m_{q_i}(m_{q_i})}{mu'} \]
where $b^{>m_q}-b^{<m_q}$ simply denotes the contribution of $q_i$ to the
beta-function coefficient above its threshold.

One expects higher loop effects to change these relations, generically by
the introduction of a power-law dependence of $\Lambda$ on
$\alpha_3(M)^{(\prime)}$ (etc.!)

\subsection{RG scale ambiguities}\label{sec:RGscale}

What is meant, in the context of a time-varying theory, by quark masses
``at $M_Z$''? Since in all likelihood the Higgs v.e.v.\ and the SU$(2)$
coupling will also be time-dependent in a unified theory, such an
expression becomes ambiguous through the change in $M_Z$. It is commonly
noted that dimensionful quantities cannot meaningfully be said to be
time-dependent, since one has no way to guarantee that units of
measurement do not also change; the presence of RG scale dependence
threatens to introduce dimensionful quantities through the back door
unless we take care in our notation.

One convenient approach is to define a reference scale which can always be
set to a constant without loss of generality; in discussing unified
theories the obvious choice is $M_U$ (or we may choose the Planck scale,
or the string scale, or the ``quantum gravity'' scale $M_\ast$ according
to the variety of theory we are considering). Then we would write in all
explicitness
\[ m_q(\mu/M_U = M_Z/M_U = f(\alpha_U)) \]
remembering that $M_Z$ will depend in the unified coupling which itself is
time-dependent; alternatively one could set a fixed ratio $\mu/M_U$ equal
to the current value $(M_Z/M_U)_{|0}\simeq 4\times 10^{-15}$ to avoid the
complication of calculating $M_Z/M_U$ as a function of $\alpha_U$. Once we
realise that the RG scale can thus be converted to a dimensionless
parameter, the ambiguity disappears.

Now we tackle the case of time-varying decoupling masses for heavy quarks
in more detail; let us take the bottom quark for definiteness. The running
mass is given at one loop by
\[ m_b(\mu) = \hat{m}_b (\alpha_3(mu))^{\gamma_1/b_1} \]
where the bottom mass runs as
\[ \frac{d}{dt} m_b = - m_b (\frac{\gamma_1}{2\pi} \alpha_3 + \cdots) \]
and $\gamma_1=4$ in QCD (note that $b_1=11-2n_f/3$ in our notation).
The decoupling mass $m_b^d$ satisfies
\[ \alpha_3^{-1}(M) + \frac{b_1}{2\pi} \ln \frac{m_b^d}{M} =
\left(\frac{m_b^d}{M}\right)^{-b_1/\gamma_1} \]
where $M$ is a constant reference mass above $m_b$ but below $m_t$ and we
have substituted the one-loop solution for $\alpha_3$ on the LHS.
Differentiating and rearranging to isolate $\Delta\ln (m_b^d/M)$, we find
\begin{equation} \label{eq:decoup}
\left(1 + \frac{\gamma_1}{2\pi}\alpha_3(m_b^d)\right)
\Delta\ln (m_b^d/M)
= \Delta\ln (\hat{m}_b/M) +
\frac{\gamma_1}{b_1}\frac{\alpha_3(m_b^d)}{\alpha_3(M)} \Delta\ln
\alpha_3(M).
\end{equation}
As remarked before, the variation in $\hat{m}_b/M$ is given just by the
change in the Higgs v.e.v.\ and in the Yukawa coupling $y_b$ at high scale
(since we are at the moment only considering the QCD contribution to the
running of quark masses, which will be the dominant effect at low
energies).

Note that the second term in brackets on the LHS of Eq.~(\ref{eq:decoup})
is formally of higher order than the first, therefore we may consistently
discard it for the heavy quarks for which $\alpha_3$ is perturbative. The
second term on the RHS is also a factor of $\alpha_3$ down compared to the
analogous expression for the variation of the QCD scale $\Lambda$ (see
Eq.~\ref{deltalnc2}), but we cannot tell without knowing the
dependence of $\hat{m}_b$ on $\alpha_U$ which term on the RHS may
dominate. If the second term on the RHS does dominate, then the changing
strength of QCD is the main cause of $m_b^d/M$ varying, in which case
quark threshold effects will be a rather small correction to the change in
$\Lambda$ and can be neglected. If the first term dominates, the main
effect is through the variation of quark masses at high energy, in which
case we can take $\Delta\ln (m_b^d/M) \simeq \Delta\ln (\hat{m}_b/M)$;
this is the relevant formula if one expects quark mass thresholds to
play a significant role.

We also note, on the subject of fermion masses, that since the fine
structure constant is defined at the scale $m_e$, a variation in $m_e/M_U$
can induce a change in $\alpha$ even if $\alpha_U$ is unchanged; this was
pointed out by Wetterich \cite{Wetterich02} in a preprint appearing after
the first version of this paper. However due to the very slow running of
$\alpha(\mu)$ one would require a rather large change in $m_e/M_U$, or in
any other charged particle threshold, to reproduce the observed variation
in alpha. This restriction, taken together with the bound on variation of
$M_p/m_e$ \cite{Ivanchik02}, imply that a sizable contribution to
varying alpha from this kind of threshold effect is unlikely
\cite{Wetterich02}.

\subsection{Dependence of $c$ on high-energy parameters}
Now we attempt to express $c^2\equiv (\hat{m}_q/\Lambda)$ as a 
function of some more fundamental parameters. Substituting for
$\hat{m}_q\equiv \hat{y} v/\sqrt{2}$, where $\hat{y}$ is the average light quark 
Yukawa coupling, modified by the appropriate function of $\alpha_3(\mu)$,
and using the expressions derived above for $\Lambda$ we find
\begin{equation} 
c^2 = {\rm const.} \times \frac{(\hat{y}_u+\hat{y}_d) v {(M')}^{6/27} 
e^{6\pi/27\alpha_3(M')}} {M'(y_c(m_c)y_b(m_b)y_t(m_t)v^3)^{2/27}}
\end{equation}
in the SM and
\begin{equation}
c^2 = {\rm const.} \times \frac{(\hat{y}_u s_\beta + \hat{y}_d c_\beta) v M^{2/3} 
e^{6\pi/27\alpha_3(M)}} {M(y_c(y_c) y_b(y_b) y_t(y_t) s^2_\beta c_\beta v^3 
m_\lambda^3 \overline{m_{\tilde{q}}^3})^{2/27}} \label{eq:csqMSSM}
\end{equation}
in the MSSM, where $s_\beta(c_\beta)=\sin(\cos)\beta$.
From now on we assume that the change in $\tan \beta$ is small and lump the
light quark Yukawas together as $\hat{y}_q$. Since the running of the up,
down and charm Yukawas as they appear in this expression ($\alpha_3$-corrected 
for the light quarks) over the range up to $M_U\sim 10^{16}\,$GeV is small, we 
set them equal to their value at unification $y_q(M_*)$. In fact, to first
approximation we neglect the nonlinear running of the $b$ and $t$ Yukawas also:
in a careful calculation one would use the semi-analytic solution for the 
top Yukawa and include the effects of the top feeding into the other masses, 
but the effects that we are neglecting by effectively taking 
$y_t(\mu)=y_t(M_*)$ for $\mu>m_t$ are likely to be sub-leading compared to, say,
a large fractional change in the Higgs v.e.v.\ or in SUSY-breaking masses.

Thus we reach the approximate expressions (which still likely contain the leading
dependence on $\alpha_U$)
\begin{eqnarray}
\Delta (\ln c^2)& =& -\frac{2\pi}{9\alpha_3(M_U)} \Delta(\ln \alpha_3(M_U)) 
+ \frac{7}{9} \Delta (\ln v/M_U) + \Delta( \ln y_{qU}) \nonumber \\
&-&\frac{2}{27} \Delta (\ln y_{cU} + \ln y_{bU} + \ln y_{tU}) 
-\frac{2}{9} \Delta (\ln m_\lambda/M_U + \ln \overline{m_{\tilde{q}}}/M_U)
\label{deltalnc2}
\end{eqnarray}
for MSSM unification and
\begin{eqnarray}
\Delta (\ln c^2)& =& -\frac{2\pi}{9\alpha_3(M_*)} \Delta(\ln \alpha_3(M_*)) 
+ \frac{7}{9} \Delta (\ln v/M_*) + \Delta( \ln y_{q*}) \nonumber \\
&-&\frac{2}{27} \Delta (\ln y_{c*} + \ln y_{b*} + \ln y_{t*}) 
\end{eqnarray}
for the SM running in low-scale models.

\section{Finding bounds on fundamental models} \label{sec:unified}
If we assume that the supersymmetric model we are referring to is associated with the low energy effective theory of some heterotic superstring or M-theory compactification the parameters $\alpha_{U}$ and $M_{U}$ are related to the size of the six dimensional manifold upon which the theory is compactified.\footnote{Although the low energy limit of M-theory is 11 dimensional, the gauge degrees of freedom, at least in the strong coupling limit of the heterotic theory, are confined to a 10 dimensional brane. The appropriate volume to consider is then that of the 6d Calabi-Yau manifold, as in heterotic string theory.} In this situation $\alpha_{U}\propto V^{-1} \propto (M_U)^{6}$ where $V$ is the volume of the compactification manifold \cite{witten}. Because of this we might expect that as one changes the size of the extra dimensions, both the GUT unification energy scale and coupling might vary together. However the effect of the changing energy scale has a negligible effect on the results because of both the 6th power in the relation between the energy scale and the coupling, and the slow running of the gauge couplings at high energies.

The supersymmetry-breaking masses entering into the expression (\ref{eq:csqMSSM}) 
depend on the mechanism of SUSY-breaking, which can either take place through
non-perturbative effects in a hidden sector (as is appropriate in models with high
fundamental scale) or perturbatively by a Scherk-Schwarz mechanism in models with
large (TeV$^{-1}$-size) extra dimensions \cite{Antoniadis,ABQ,Delgado}.\footnote{If the fundamental scale is intermediate ($10^{11}$--$10^{13}\,$GeV) 
then SUSY-breaking may occur directly at tree level in a non-supersymmetric 
hidden sector and be gravitationally-mediated, a possibility recently realized 
in certain D-brane models \cite{Quevedo}.} Taking the scenario of gaugino 
condensation in a hidden sector and string moduli mediation as a benchmark, 
the functional dependence on the unified coupling is
\begin{equation}
\tilde{m} \propto m_{3/2} \propto \Lambda_H^3 M_P^{-2} 
\end{equation}
where $M_P$ is the reduced Planck mass $1/\sqrt{8\pi G_N}$ and the hidden sector
confinement scale is $\Lambda_H^3 \sim M_U^3 e^{-6\pi/b_H \alpha_H(M_U)}$.
Thus the fractional change expected is related as 
\begin{equation}
\Delta(\ln \tilde{m}/M_U) \simeq 2 \Delta\log(M_U/M_P) -\frac{6\pi}{b_H\alpha_U}
\Delta(\ln \alpha_U) 
\end{equation}
where the hidden sector gauge coupling is taken also to unify at $M_U$. 
Using the relation $M_U\sim M_s \propto g_U M_P$ from heterotic
string theory, and imposing the correct magnitude of 
$e^{-6\pi/b_H \alpha_H(M_U)}\sim \tilde{m}/M_P \simeq 10^{-15}$ we find 
\begin{equation}
\Delta(\ln \tilde{m}/M_U) \simeq (2 + 34.5) \Delta(\ln \alpha_U) 
\end{equation}
where the factor $34.5 \simeq 15\ln 10$ is universal to models of 
SUSY-breaking by strong coupling effects in a hidden sector mediated by
non-renormalisable operators (as also found in \cite{LSS}).

The quark Yukawa couplings gain a universal factor of 
$(S+\bar{S})^{-1/2}\sim g_U$ from the normalisation of the superpotential
in going from SUGRA to softly-broken global SUSY, thus in particular
$y_{tU}$ will vary proportional to $\alpha_U^{1/2}$ which will be of importance
for electroweak breaking. The light quark Yukawas are extremely 
model-dependent, but estimates can be made in some classes of
unified models. The most common mechanism employed to generate small Yukawa
couplings is the Froggatt-Nielsen picture in which 
effective couplings arise from non-renormalisable operators when scalars charged
under a U$(1)$ group get v.e.v.'s $\langle X \rangle$. In fact this picture
finds a natural embedding in the heterotic string where the group is now
``pseudo-anomalous'' and is broken near the string scale due to a nonzero
Fayet-Iliopoulos term at one loop destabilising the symmetric ($X=0$) vacuum.
The fact that this is a one-loop effect tells us the dependence of the v.e.v.
on the unified coupling as $\langle X \rangle/M_P \propto g_U$ (see {\em e.g.}\/\
\cite{Giedt}), thus the small Yukawas are generated schematically as
\begin{equation}
y_{iU} \sim \langle X \rangle^{Q_i} \propto g_U^{Q_i}
\end{equation}
where the quark $q_i$ has a U$(1)$ charge $Q_i$ of opposite sign to $Q_X$.
Thus with knowledge of the charges, which also follow phenomenologically if 
we know the constant of proportionality of $\langle X \rangle/M_P = k g_U$ and 
demand the correct values of $y_{iU}$, one easily finds the variation in 
$y_{iU}$ as a function of $\Delta (\ln \alpha_U)$.

Finally, as is well known the standard scenario of radiative electroweak 
sym\-metry-breaking is sensitive to the soft masses $\tilde{m}$ and the top
Yukawa $y_{tU}$ (for a review, see \cite{IbanezRoss}). While the full RG 
equations for the top Yukawa and the relevant soft masses are complicated,
the leading dependence on the input quantities was already considered in
\cite{RobertsRoss} in which the following estimate was made for the 
sensitivity of the $W$ mass to $y_t$
\begin{eqnarray}
M_W(\mu<Q_0) &\simeq& \frac{g_2(\mu) v^2}{4} \propto y_t m_{\tilde{t}} 
\ln(\mu^2/Q_0^2) \nonumber \\
\rightarrow \Delta (\ln M_W(\mu)/M_U) &\approx& \Delta (\ln \tilde{m}/M_U)
+ \Delta (\ln y_t) \frac{\ln M_U/\mu}{\ln Q_0/\mu}
\end{eqnarray}
where $Q_0$ is the scale at which the up-type Higgs mass-squared crosses to
negative values, given by $M_U e^{-k/y_t^2}$, where $k$ is a ratio of 
SUSY-breaking masses at the scale $M_U$ and we choose a fixed renormalisation 
scale $\mu\simeq M_{W|0}$ since (analogously to the case of heavy quark 
thresholds) the running of $g_2(\mu)$ is slow. Then we impose 
$M_U/\mu \simeq 10^{14}$: one finds from a more careful analysis 
\cite{RobertsRoss} that $Q_0/\mu$ tends to be of order 10, in which case the
enhancement factor $\ln (M_U/\mu)/\ln (Q_0/\mu)$ is of order 15 (although
this also depends strongly on the pattern of Higgs mass parameters at 
unification). 

Hence in addition to the strong SUSY-breaking mass effect previously discussed, 
there is a milder dependence of $M_W$ or $v$ through $y_t\propto \alpha_U^{1/2}$ 
and the overall effect on the Higgs v.e.v. may well be the largest contribution 
to any low-energy physics that depends on quark masses, even relative to the 
exponential dependence of $\Lambda$ on the high-energy coupling strength in QCD. 
The reason for this is subtle: while the 
beta-function coefficients giving the running of the hidden sector gauge 
coupling are assumed constant, thus the hidden sector scale $\Lambda_H^3$ 
``feels'' the whole of the hierarchy between $M_U$ and $M_W$, the presence of 
thresholds at low energy ($\ll M_U$) in QCD means that $\Lambda_{\rm QCD}$ feels 
mainly the running at low energies ($<\tilde{m}$) thus the sensitivity to varying 
the coupling at energies above the thresholds is less. If there were also 
thresholds for charged states in the hidden sector at masses $\Lambda_H 
\lesssim M_H \ll M_U$, 
%but above $\Lambda_H$ 
the leading dependence of SUSY-breaking on the unified coupling would be 
smaller, however such a case is atypical \cite{LustT+deCCM91} for the 
vector-like matter representations usually considered.

However, since SUSY-breaking and electroweak breaking are highly 
model-dependent, we will not go through the resulting bounds in detail. 
Other types of model are likely to give completely different results,
for example Scherk-Schwarz SUSY-breaking, where the electroweak and
SUSY-breaking mass scales vary inversely with the radius of an extra dimension, 
as do the 4-d gauge couplings (more precisely, $\alpha_i\propto R^{-1}$) and so
the fractional variation in $v$ would likely be of the same order as that
in $\alpha_U$ (rather than maybe 50 times larger).

\section{Discussion and Constraints}
In drawing our conclusions about the bounds on the variation of $\alpha_U$, 
we will simply for the sake of an easy comparison set to zero all variation 
in quark masses and other thresholds, from no matter what source, leaving
the effect of $\alpha_3$ on $\Lambda$ as the only varying quantity. 
This is not intended as a realistic treatment of unified theories but only 
as an illustration of one effect, which can be systematically extended to 
whatever unified theory by going back to the bounds on 
$c\equiv (\hat{m}_q \Lambda)^{1/2}$ and substituting for the 
expressions we derived in Sections \ref{sec:thresh} and \ref{sec:unified}.
With this caveat we proceed to discuss the bounds from the di-nucleon systems.

\subsection{The deuteron}

A 69\% increase in $c^{-2}=\Lambda/(m_u+m_d)$ will destabilise the deuteron. On the assumption of fixed quark masses, Fig.~\ref{variation} tells us that this corresponds to a 3\% increase in the gauge coupling at the scale of unification for a conventional SUSY GUT. This creates problems for cosmological models where the gauge couplings vary significantly over cosmological time scales \cite{brax}, since no such variation can have occurred at any time since nucleosynthesis.

For a model with a TeV scale GUT, $\alpha_3$ cannot have increased more than 8\% at any time since nucleosynthesis, since this would also lead to the deuteron becoming unbound.

Recent observations suggest a {\em negative}\/ variation in the electromagnetic fine structure constant at high redshift \cite{webb}, which would, if one assumes high scale gauge unification and consequently a corresponding negative change in $\alpha_3$, increase the binding energy of the deuteron. Unless the field responsible for the value of the gauge coupling is frozen by the Hubble expansion \cite{DvaliZ} one might expect this variation to be much larger in the early universe (see {\em e.g.}\/\ \cite{barrow}). This might result in the deuteron having been more resilient to photo-dissociation by the decay products of massive relic particles \cite{ellis}. This possibility would be beneficial to models where the decay of gravitinos can cause problems for the light element abundances of nucleosynthesis. 

On the other hand, such an increase in the binding energy would allow helium to form at higher temperatures, leading to an increase in the primordial helium abundance. This effect, neglected in \cite{olive} and \cite{LSS}, would be superimposed on other effects due to changing couplings, {\em i.e.}\/\ a change in freezeout temperature and neutron lifetime. One can make a rough estimate of the sensitivity of the helium abundance to the deuteron binding energy in the following way. The ratio of number densities of the neutron to the proton $n/p$ at the time of helium formation determines the primordial abundance of helium, since 99.99\% of the neutrons go on to form helium. This is related to the initial ratio of neutron to proton number density at the time of weak interaction freezeout $n_0/p_0$, via the decay of neutrons into protons, by
\begin{equation}
\frac{n}{p}=\frac{e^{-t_{He}/\tau}}{p_0/n_0 +1-e^{-t_{He}/\tau}}
\end{equation}
where $t_{He}$ is the time at which helium production takes place and $\tau$ is the neutron lifetime (currently measured to be $887\,s$). In this formula, weak interactions are taken to freeze out at $t=0$.  We have also assumed that all helium production occurs instantaneously at $T_{He}$. Although still a subject of some controversy, the orthodox view of light element abundances determines the initial mass fraction to an accuracy of $\pm 5\%$ is the ratio $n/p$ \cite{pagel}.  Translating this into a constraint upon the variation of $T_{He}$ we obtain
\begin{equation}  
-0.20\le\frac{\Delta T_{He}}{T_{He}}\le +0.13
\end{equation}
It is reasonable to suggest that $T_{He}$ is proportional to the binding energy of the deuteron, since only below a certain temperature set by that binding energy can helium formation proceed unimpeded by the photo-dissociation of the deuteron.  This is because the reactions which form helium all rely upon deuterium as an intermediate building block :-
\begin{eqnarray}
{\rm{H}}^{2}({\rm{H}}^{2},n){\rm{He}}^{3}({\rm{H}}^{2},p){\rm{He}}^{4}& & \nonumber\\
{\rm{H}}^{2}({\rm{H}}^{2},p){\rm{H}}^{3}({\rm{H}}^{2},n){\rm{He}}^{4}& & \nonumber\\
{\rm{H^{2}}}({\rm{H^{2}}},\gamma){\rm{He}}^{4}& &.
\end{eqnarray}
Using Eq.~(\ref{cotcot}) we can find out how the parameter $\beta$ changes for different values of the parameter $c$.  In order to translate this into the binding energy $E_B=\beta V_{sq}$ it is necessary to take into account the scaling of the depth of the square well $V_{sq}\propto m_{\pi}/c^2\propto m_q^{\frac{3}{2}}\Lambda^{-\frac{1}{2}}$ so the result is not in terms of the dimensionless parameter $c$.  Therefore we take the case where the variation in the quark Yukawa couplings is not significant and simply consider the variation in $\Lambda$. The result is
\begin{equation}
-0.04 \le \frac{\Delta \Lambda}{\Lambda}\le +0.04
\end{equation}
which relates into constraints on the gauge couplings at 1 TeV in non-SUSY models or $10^{16}$ GeV for GUT models of 
\begin{eqnarray}
-0.005\le\left.\frac{\Delta\alpha_3}{\alpha_3}\right|_{1\,{\rm TeV,\ no\ SUSY}}\le 0.005\nonumber\\
-0.0023\le\left.\frac{\Delta\alpha_3}{\alpha_3}\right|_{10^{16}\,{\rm GeV,\ SUSY}}\le 0.0023
\end{eqnarray}
which is obviously very restrictive.  In order to investigate this further it would be necessary to obtain a more detailed model for the deuteron binding energy and the various binding energies, cross sections and decay rates in nucleosynthesis.  This effect then has to be added to the effect of varying neutron-proton mass difference and the effect of gauge coupling variation on weak coupling $\cite{olive}$.  However the point remains that in general a variation in the gauge coupling at nucleosynthesis would have a very large effect upon nuclear binding energies, and consequently on the primordial abundance of light elements.  This would create problems for models such as $\cite{melch}$ where a change in $\alpha$ of order 1\% at nucleosynthesis is considered.

If one decreases $\Lambda$, the binding energy of the deuteron increases, therefore helium is produced earlier when there are more neutrons and the helium abundance goes up.  At the same time, a decrease in $\Lambda$ decreases the neutron-proton mass difference, which increases the number of neutrons and therefore the helium abundance, so one would expect that two effects should work together, implying a yet more restrictive bound.

\subsection{Implications of a stable di-neutron or di-proton}
A negative change in the gauge coupling at high redshift would decrease the Coulomb repulsion and increase the nuclear forces in the di-proton system.  The di-proton and di-neutron systems only become stable if $\Lambda$ 
decreases to around one tenth of its present day value.  This corresponds to a 
10\% decrease in the unified gauge coupling for a high scale model and a 
25\% decrease for a TeV GUT.  
The stability of the di-proton would have a catastrophic effect upon the lifetime of stars as it would provide a rapid channel for hydrogen fusion \cite{dyson}. However the constraints on the gauge coupling variation in the matter dominated epoch are several orders of magnitude too small for this to occur \cite{webb}. The presence of the di-proton would also be disastrous for nucleosynthesis and might eliminate all the hydrogen in the universe.  However, the large negative change in the gauge couplings required indicates that this is not a particularly strong constraint.

It is not immediately obvious how dangerous the stability of the di-neutron would be to nucleosynthesis, since the neutrons would probably still end up in Helium atoms.  Still, we can safely say that the variation in the gauge coupling required for the di-proton and di-neutron to become stable is much larger than the typical orders of magnitudes being considered at the moment.

\section{Conclusions}
We have shown in this paper that a 3\% increase in the QCD coupling constant $\alpha_3$ at the GUT scale would result in the deuteron becoming unbound. The deuteron binding depends only on nuclear forces, so this conclusion cannot be escaped by considering the variation of more than one gauge coupling at once. Only negative variations at the level of 10\% could bind the di-proton and the di-neutron.

We have developed formalisms which enable one to calculate the variation in low energy parameters as a function of the variation of gauge and Yukawa couplings in the underlying theory: in many models one expects the variation in the electroweak and SUSY-breaking sectors to be the dominant effect at low energies, but the model-dependent nature of such effects means that no firm conclusions can yet be reached.

We have also considered the effect of variations in the binding energy of the deuteron on the time at which helium formation occurs, and consequently on the helium abundance. This effect is complementary to the other effects on nucleosynthesis due to variation of gauge couplings, but on its own it constrains variation in $\alpha_3$ at nucleosynthesis to within 0.25 \%.  

\section*{Acknowledgements}
We would like to thank David Bailin, John Barrow, Ed Copeland, Marty Einhorn, Gordy Kane, Bernard Pagel and Lian-Tao Wang for valuable conversations.


\begin{thebibliography}{99}
\bibitem{wood} This method for calculating the stability is the work of David M.\ Wood, who developed it to investigate the stability of central Yukawa potentials.
\end{thebibliography}
\end{document}